\documentclass[pra,aps,superscriptaddress,balancelastpage,onecolumn]{revtex4}
\usepackage{amsmath}
\usepackage{graphicx}
\usepackage{amsfonts}
\usepackage{amssymb}
\usepackage{soul,xcolor}
\usepackage{lmodern}
\usepackage[utf8]{inputenc}
\usepackage[T1]{fontenc}
\providecommand{\U}[1]{\protect\rule{.1in}{.1in}}
\providecommand{\U}[1]{\protect\rule{.1in}{.1in}}

\begin{document}
\setstcolor{red}
\title{Nonreciprocal Multipartite Entanglement in a two-cavity
magnomechanical system }
\author{Rizwan Ahmed}
\affiliation{Physics Division, Pakistan Institute of Nuclear Science and Technology
(PINSTECH), P. O. Nilore, Islamabad 45650, Pakistan}
\author{Hazrat Ali}
\affiliation{Department of Physics, Abbottabad University of Science and Technology, P.O. Box 22500 Havellian KP, Pakistan}
\author{Aamir Shehzad}
\affiliation{Department of Physics, Government College University, Allama Iqbal Road,
Faisalabad 38000, Pakistan.}
\author{S K Singh}
\affiliation{Process Systems Engineering Centre (PROSPECT), Research Institute of Sustainable Environment (RISE),  School of Chemical and Energy Engineering, Universiti Teknologi Malaysia, Johor Bahru 81310, Malaysia}
\author{Amjad Sohail}
\email{amjadsohail@gcuf.edu.pk;amjadss@ifi.unicamp.br}
\affiliation{Department of Physics, Government College University, Allama Iqbal Road,
Faisalabad 38000, Pakistan.}
\affiliation{Instituto de F\'{\i}sica Gleb Wataghin, Universidade Estadual de Campinas, Campinas, SP, Brazil}
\author{Marcos C\'esar de Oliveira}
\email{marcos@ifi.unicamp.br}
\affiliation{Instituto de F\'{\i}sica Gleb Wataghin, Universidade Estadual de Campinas, Campinas, SP, Brazil}

\begin{abstract}
We propose a theoretical scheme for the generation of nonreciprocal multipartite entanglement in a two-mode cavity magnomechanical system, consisting of two cross-microwave (MW) cavities having a yttrium iron garnet (YIG) sphere, which is coupled through magnetic dipole interaction. Our results show that the self-Kerr effect of magnon can significantly enhance multipartite entanglement, which turns out to be nonreciprocal when the magnetic field is tuned along the crystallographic axis [110]. This is due to the frequency shift on the magnons (YIG sphere), which depends on the direction of the magnetic field. Interestingly, the degree of nonreciprocity of bipartite entanglements depends upon a careful optimal choice of system parameters like normalized cavity detunings, bipartite nonlinear index $\Delta E_{K}$, self-Kerr coefficient, and effective magnomechanical coupling rate $G$. In addition to bipartite entanglement, we also present the idea of a bidirectional contrast ratio, which quantifies the nonreciprocity in tripartite entanglements. Our present theoretical proposal for nonreciprocity in multipartite entanglement may find applications in diverse engineering nonreciprocal devices.

\end{abstract}

\maketitle

\section{Introduction}
Since the early days of quantum theory, entanglement at the macroscopic scale has been a significant topic of discussion. Today, it is recognized as a fundamental resource at the core of modern quantum information processing \cite{schrodinger,epr,horodeki}. Understanding entanglement is essential to grasp the boundary between classical and quantum physics \cite{class}, and it has numerous technological applications, such as quantum sensing \cite{app,app2}, quantum networks, and multi-tasking in quantum information processing \cite{network,network2}. Entanglement has been successfully generated in various systems. Of particular relevance to our discussion is the theoretical and experimental demonstration that entanglement can be produced in nonlinear quantum systems, specifically cavity optomechanical systems \cite{opto,opto1,opto2,opto3,opto4} and magnomechanical systems \cite{mag,mag1,mag2,mag3,mag4}.

Recently, nonreciprocal entanglement has gained considerable attention in macroscopic quantum systems due to a wide range of applications in cloaking (invisible sensing) and noise-free information processing \cite{non}. This is because entanglement can be well protected by Lorentz reciprocity breaking \cite{lorentz}. Several nonreciprocal devices have been proposed and realized in cavity optomechanical systems \cite{prop,device,device2,device3}. Furthermore, Lorentz symmetry can be broken by introducing the Sagnac effect \cite{sagnac,sagnac2}, which results in nonreciprocal entanglement \cite{Nnent}. We are particularly interested in electromagnetic reciprocity, where using magneto-optical materials leads to the breaking of Lorentz symmetry. Although the physical realization of these devices is difficult due to highly susceptible external magnetic field interference \cite{diff}, there are various systems such as optomechanics \cite{op}, non-Hermitian optics \cite{nH}, nonlinear optics \cite{NL} and atomic-gases-based systems \cite{gas} that exhibit nonreciprocal features. It is important to mention that most of these previous studies mainly addressed the classical regime focused on the transmission rate non-reciprocity. However, recently, quantum devices have been explored based on the Fizeau light-dragging effect \cite{fiz}, nonreciprocal photon blockade \cite{block}, and backscattering immune optomechanical entanglement \cite{imm}. Furthermore, nonreciprocal magnon and phonon lasers have been proposed, which use the similar Fizeau light-dragging effect \cite{YJX,YJX2}.

Cavity magnomechanical systems may constitute a valuable platform for the investigation of nonreciprocal multipartite entanglement, as they have drawn considerable attention due to their high spin density and lower collective losses \cite{losss}. Typically, cavity magnomechanical systems (CMMS) contain magnetically ordered materials such as yttrium-iron-garnet (YIG), which can be strongly coupled to microwave (MW) cavity fields. In addition to magnetic interactions with magnons, magnetic modes in CMMS can also have magnetostrictive interactions, leading to the coupling of magnon modes with mechanical/phonon modes \cite{strict}. These systems allow for the investigation of several quantum phenomena at the microscopic level, such as dark modes \cite{dark}, entanglement \cite{mag,mag1,mag2,mag3,mag4}, magnon blockade \cite{Ir1,Ir4}, and nonlinear dynamics, including collapse and revival mechanisms, have shown remarkable success in magnon-cavity interactions \cite{IR5}, perfect absorption \cite{abs}, unconventional magnon excitations \cite{uncon}, and quantum steering \cite{mag2}. Furthermore, many of the recent studies discussed the effects of Kerr nonlinear medium on the system dynamics. These include the interaction of the cavity field with three-level atoms inside a f-deformed cavity \cite{Ir6}. In other interesting papers, quantum statistics and entanglement dynamics have been extensively discussed by employing a ferrimagnetic medium inside an optomechanical cavity \cite{IR2,IR3}. Apart from a plethora of theoretical studies and proposals in cavity magnonic systems, various attempts at experimental implementations are also realized for many practical applications \cite{exp,exp1,exp2,exp3}.

YIG spheres, with a typical size of $0.1$mm, provide an alternative method to investigate macroscopic quantum processes such as entanglement and squeezing. Research has shown that the Kerr nonlinearity, which can arise from magnetocrystalline anisotropy, enhances the degree of entanglement and nonclassicality \cite{YPU}. For example, the Kerr effect helps to enhance steady-state bipartite and tripartite entanglement in an optomagnonical system \cite{JJC,yang2020entanglement}.
Previous approaches to analyzing nonreciprocal entanglement have primarily focused on the Sagnac effect, which causes a positive or negative shift in the cavity resonance frequency based on the direction of the driving force. On the other hand, the Kerr effect in magnomechanical systems can
also induce positive or negative frequency shifts depending on the
direction of the magnetic field. Unlike the Sagnac effect, the magnon Kerr
effect generates a two-magnon supplemental effect, which enhances the optimum
values of all entanglements in our present configuration \cite{MYJOS,MYJOS1,IR3}%
. As a result, both bipartite and tripartite entanglements are
increased/shifted when compared to the scenario without the Kerr effect.
Tuning the aligned magnetic field along the crystallographic axis [100] or
[110] allows for nonreciprocally formed entanglements. In the present
manuscript, we aim to study the nonreciprocal multipartite
entanglement in a cavity magnomechanical system (See Fig.1). Furthermore, it was shown that the Kerr effect modifies the magnon number through frequency shift in a way that is adequate to strengthen bipartite entanglements between various bipartite and tripartite entanglement.

The paper is organized as follows. Section II introduces the basic magnomechanical system under consideration and the corresponding Hamiltonian. Additionally, quantum Langevin equations are derived using this Hamiltonian. In Section III, the basic entanglement quantification scheme based on logarithmic negativity is discussed. Next, in Section IV, results and discussion are presented for different system parameters. Finally, Section V concludes the paper.

\begin{figure}[tbp]
\centering
\includegraphics[width=0.98\columnwidth,height=4in]{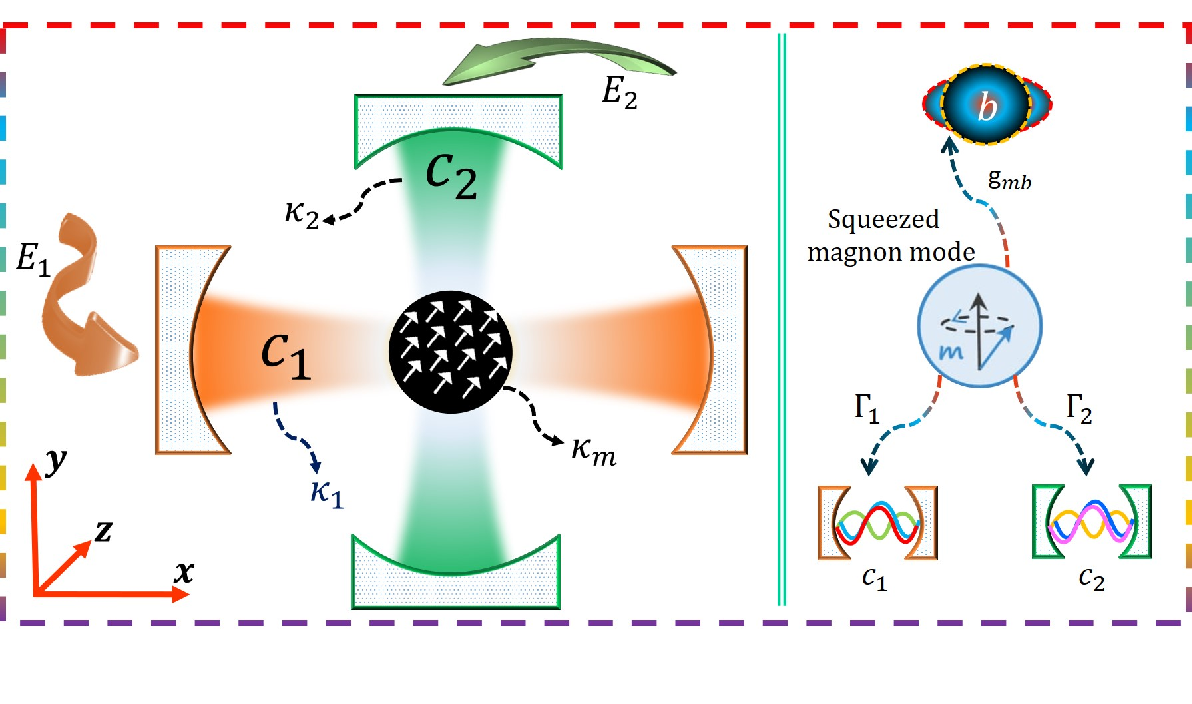} \centering
\caption{Schematic illustration of our proposed system where a YIG sphere (with Kerr-nonlinearity) simultaneously couples with two driven MW cavity modes, through magnetic dipole interaction.
The YIG sphere is positioned along its crystallographic axis ([100] or [110]) in a static magnetic field. In
addition, vibrational motion is considered as a  phonon mode, which couples with the magnon modes owing to a magnetostrictive effect.
}
\end{figure}
\section{THEORETICAL MODEL AND HAMILTONIAN OF THE SYSTEM}
We consider a magnomechanical system with a YIG sphere, supporting a Kittel mode, placed inside dual cross-shaped MW cavities (see Fig. 1). Magnons are quasiparticles that show the collective excitation of spins within a ferrimagnet \cite{1,mag}. The magnons and the cavity mode are coupled via magnetic dipole interaction. In our present magnomechanical system, the magnetic fields of the two cavity modes are along the $x$ and $y$ directions. Additionally, the external magnetic field is oriented in the $z$-direction. The choice of the Cartesian system shown in Fig. 1 is completely arbitrary, and we can choose any orientation keeping the magnetic field perpendicular to the cavity fields. In addition, the coupling between the phonon and magnon modes is mediated by the magnetostrictive interaction caused by the YIG sphere's geometrical deformity \cite{Kittel,Zha}. By employing the rotating wave approximation at the driving field frequency $\omega _{0}$, the Hamiltonian of the dual-cavity magnomechanical system can then be formulated as
\begin{eqnarray}
H &=&\sum_{j=1}^{2}\Delta _{j}c_{j}c_{j}^{\dag }+\Delta _{m_{0}}m^{\dag }m+%
\frac{\omega _{b}}{2}\left( q^{2}+p^{2}\right)+\Gamma _{j}(c_{j}m^{\dag }+c_{j}^{\dag }m)+g_{mb}m^{\dag }mq \notag \\
&&+\mathcal{K}_{r}m^{\dag }mm^{\dag }m +i\Omega \left( m^{\dag }-m\right) +iE_{j}(c_{j}^{\dag }-c_{j}),
\label{HML}
\end{eqnarray}%
where $\Delta _{m_{0}}=\omega _{m}-\omega _{0}$ and $\Delta _{j}=\omega
_{j}-\omega _{0}$ ($j=1,2$). Here, $\omega _{m}$ ($\omega _{j}$) is the
resonance frequencies of the magnon ($j$th cavity) mode while $\omega _{0}$
is the frequency of the drive magnetic field. The first term in Eq. (\ref%
{HML}) reflects the free Hamiltonian of two cavity modes, where $c_{j}^{\dag }$
and $c_{j}$ ($j=1,2$) are the creation and annihilation operators for the two cavity modes. Similarly, the second term is the free Hamiltonian of the magnon mode, where $m^{\dag }$ and $m$ are the creation and annihilation operators of the magnon mode. It is worth mentioning that the operators $m^\dag$ and $m$ are the bosonic field operators for magnons and do not represent the applied external magnetic field. The
third term is the free Hamiltonian of the phonon mode, where $%
q $ and $p$ are the dimensionless position and momentum quadratures of the
phonon mode with vibrational frequency $\omega _{b}$.
It is worth noting
that the magnon frequency can be adjusted through the biased magnetic
field, $H$, via $\omega _{m}=\gamma _{G}H$, where $\gamma _{G}$ is denoted the
gyromagnetic ratio. The fourth term represents the interaction between the $%
j $th cavity and the magnon modes with optomagnonical coupling strength $%
\Gamma _{j}$ \cite{AORH}, which is given by
\begin{equation}
\Gamma _{j}=\mathcal{V}\frac{c}{n_{r}}\sqrt{\frac{2}{\rho _{spin}V_{m}}},
\end{equation}%
where $\mathcal{V}$ is the Verdet constant, $V_{m}=\frac{4\pi r^{3}}{3}$ is the volume, $\rho _{spin}$ is the spin density, and $n_{r}$ is the refractive index for the YIG sphere.
The fifth term represents the interaction between the magnon and the mechanical modes, with $g_{mb}$ being the strength of the magnetostrictive effect-induced magnonmechanical coupling.
The sixth term in Eq. (\ref{HML}) represents the magnon self-Kerr nonlinear
term, with the self-Kerr coefficient $\mathcal{K}_{r} $, resulting in magnon
squeezing, which is precisely proportional to the quadratic magnon field
operator. Here, $\mathcal{K}_{r}=\mu_{0}k_{an}\gamma_{G}^{2}[M^{2}V_{yig}]^{-1}$,
in which $\mu_{0}$ is the magnetic permeability of free space, $k_{an}$ represents the first-order anisotropy constant of YIG sphere, and $M$ indicates the saturation magnetization. Furthermore, $\mathcal{K}_{r}$ becomes prominent for small-sized YIG spheres. In addition, the magnon self-Kerr coefficient $\mathcal{K}_{r} $ can be positive (negative) depending upon the magnetic field alignment along the crystallographic axis [100] ([110]). Furthermore, it can be tuned from $0.05$
nHz to $100$ nHz depending on the diameter of the YIG sphere, which ranges
from $1$ mm to $100\mu$m. The Rabi frequency $\Omega=\frac{5}{4}\gamma_{G} \sqrt{N}H_d$ indicates the strength of the coupling between the driving field of the microwave and the magnon, where $N=\rho V_{yig}$ stands for the YIG crystal's total spin number, with $\rho$ ($V_{yig}$) being the density (volume) of the YIG sphere, and $H_d$ the drive magnetic field's amplitude. Finally, $E_j=\sqrt{2k_jP_j/\hbar\omega_{L_j}}$ is the coupling strength between the driving laser field and the cavity photon, where $k_j$ is the decay rate of the cavity field and $P_j(\omega_{L_j})$ is the power (frequency) of the input laser field.
\section{THE QUANTUM DYNAMICS AND MULTIPARTITE ENTANGLEMENT}
In this section, we calculate the quantum dynamics of a two-cavity
magnomechanical system employing the standard Langevin approach. Taking
into account the dissipation-fluctuation process, the quantum Langevin
equations for the two-cavity magnomechanical system, are as follows
\begin{eqnarray}
\dot{q} &=&\omega _{b}p,  \notag  \label{LG} \\
\dot{p} &=&-\omega _{b}q-g_{mb}m^{\dag }m-\gamma _{b}p+\xi ,  \notag \\
\dot{m} &=&-i\Delta _{m}^{0}m-i\sum_{j=1}^{2}\Gamma
_{j}c_{j}-ig_{mb}mq+\Omega -2i\mathcal{K}_{r}m^{\dag }mm-\kappa _{m}m+\sqrt{2\kappa _{m}}m^{in},
\notag \\
\dot{c}_{j} &=&-i\Delta _{j}c_{j}-i\Gamma _{j}m-\kappa _{j}c_{j}+E_{j}+\sqrt{%
2\kappa _{c}}c_{j}^{in},  \label{LEEE}
\end{eqnarray}%
where $%
\gamma _{b}$, $k_{m}$ and $k_{j}$ are the decay rates of the phonon, magnon, and cavity modes, respectively, while their corresponding input noise
operators are $\xi $, $m^{in}$ and $c_{j}^{in}$. The noise operators for the cavity and magnon modes must fulfill the following correlation functions:$\left\langle \Lambda ^{in\dag }(t)\Lambda ^{in}(t^{\prime
})\right\rangle =n_{l}(\omega _{l})\delta (t-t^{\prime })$, and $%
\left\langle \Lambda ^{in}(t)\Lambda ^{in\dag }(t^{\prime })\right\rangle
=[n_{l}(\omega _{l})+1]\delta (t-t^{\prime })$, where $\Lambda
=m,c_{1},c_{2} $, $l=m,1,2$, and $n_{l}(\omega _{l})=[\exp (\frac{\hbar
\omega _{l}}{k_{B}T})-1]^{-1}$ is the thermal magnon(photon) number
related to magnon(cavity) modes, where $T$ the temperature and $k_{B}$
denotes the Boltzmann constant \cite{Gardiner}. Furthermore, the correlation functions
corresponding to the phonon damping rate are given by $\left\langle \xi (t^{\prime
})\xi (t)\right\rangle +\left\langle \xi(t)\xi (t^{\prime })\right\rangle
/2=\gamma _{b}[2n_{b}(\omega_{b})+1]\delta (t-t^{\prime })$, where $n_{b}$ is the average number of phonon number at thermal equilibrium, and is given by $n_{b}(\omega _{b})=[\exp (\frac{\hbar \omega _{b}}{k_{B}T})-1]^{-1}$.

Below we adopt the standard linearization method to derive the linearized
Langevin equation. We rewrite each operator of the Eq.(\ref{LEEE}) as the
sum of the mean value and the fluctuation part,
\cite{Shl1,Shl2} i.e., $\mathcal{Q}=\left\langle \mathcal{Q}%
\right\rangle +\delta \mathcal{Q},(\mathcal{Q}=p,q,c_{k},m)$.
Therefore, we obtain the following mean value of the operators:
\begin{eqnarray}
p_{s} &=&0,\notag \\
q_{s}&=&\frac{-g_{mb}}{\omega _{b}}\left\vert
m_{s}\right\vert ^{2}, \notag \\
c_{j,s} &=&\frac{E_{j}-i\Gamma _{j}m_{s}}{\kappa {j}+i\Delta _{j}},  \notag
\\
m_{s} &=&\frac{-i\Gamma _{1}E_{1}\alpha _{2}-i\Gamma _{2}E_{2}\alpha
_{1}+\Omega \alpha _{1}\alpha _{2}}{\alpha _{1}\alpha _{2}\alpha _{m}+\Gamma
_{1}^{2}\alpha _{2}+\Gamma _{2}^{2}\alpha _{1}},  \label{MAV}
\end{eqnarray}%
where $\tilde{\Delta}_{m}=\Delta _{m}+\Delta K$, with $\Delta _{m}=\Delta
_{m}^{0}+g_{mb}\left\langle q\right\rangle $ and $\Delta K=2\mathcal{K}%
_{r}\left\vert m_{s}\right\vert ^{2}$ and $\alpha _{f}=k_{f}+i\Delta _{f}$ $%
(f=1,2,m)$. As $\mathcal{K}_{r}$ might be positive (negative), resulting in $%
\Delta K>0$ ($\Delta K<0$). Therefore, the steady-state magnon number can be
sufficiently tuned via $\Delta K$, resulting in a nonreciprocal mean magnon
number. For $|\Delta _{m}|$,$|\Delta _{j}|\gg \kappa _{m}$,$%
\kappa _{j}$, Eq. (\ref{MAV}) yields
\begin{equation}
m_{s}=i\left[ \frac{\Delta _{2}\Gamma _{1}E_{1}+\Delta _{1}\Gamma
_{2}E_{2}-\Omega \Delta _{1}\Delta _{2}}{\Delta _{1}\Delta _{2}\Delta
_{m}-\Delta _{1}\Gamma _{2}^{2}-\Delta _{2}\Gamma _{1}^{2}}\right] ,
\end{equation}%
which is a pure imaginary number. Neglecting high-order fluctuation terms,
the linearized equations can be extracted as
\begin{eqnarray}
\delta \dot{q} &=&\omega _{b}\delta p,  \notag \\
\delta \dot{p} &=&-\omega _{b}\delta q-\gamma _{b}\delta p-g_{mb}\left(
\left\langle m\right\rangle \delta m^{\dag }+\left\langle m\right\rangle
^{\ast }\delta m\right) +\xi ,  \notag \\
\delta \dot{m} &=&-(i\bar{\Delta}_{m}+\kappa _{m})\delta
m-i\sum_{k=1}^{2}\Gamma _{k}\delta c_{k}-i\Delta _{\mathcal{K}_{2}}\delta
m^{\dag } -ig_{mb}\left\langle m\right\rangle \delta q+\sqrt{2\kappa _{m}}m^{in},
\notag \\
\delta \dot{c}_{k} &=&-\left( i\Delta _{k}+\kappa _{k}\right) \delta
c_{k}-i\Gamma _{k}\delta m+\sqrt{2\kappa _{a}}c_{k}^{in},
\end{eqnarray}%
where $\bar{\Delta}_{m}=\Delta _{m}+\Delta _{\mathcal{K}_{1}}$ s the modified detuning, which includes an additional contribution of frequency shift by magnon squeezing, i.e., $\Delta _{\mathcal{K}_{1}}=4\mathcal{K}%
_{r}\left\vert m_{s}\right\vert ^{2}=2\Delta K$. In addition, $\Delta _{%
\mathcal{K}_{2}}=2\mathcal{K}_{r}m_{s}^{2}=-2\mathcal{K}_{r}\left\vert
m_{s}\right\vert ^{2}=-\Delta K$, ($ \because m_{s}^{*}=-m_{s}$).
The quadratures correspond to the magnon and cavity
fluctuation operators can be defined as $\delta x=\frac{1%
}{\sqrt{2}}(\delta m^{\dag }+\delta m)$, $\delta y=\frac{i}{\sqrt{2}}(\delta
m^{\dag }-\delta m)$, $\delta X_{j}=\frac{1}{\sqrt{2}}(\delta c_{j}^{\dag
}+\delta c_{j})$ and $\delta Y_{k}=\frac{i}{\sqrt{2}}(\delta c_{j}^{\dag
}-\delta c_{j})$. We can write the fluctuation equations, after incorporating these defined quadratures, in compact matrix for as,
\begin{eqnarray}
\dot{\digamma}(t)=\mathcal{M}\digamma (t)+\mathcal{N}(t),
\end{eqnarray}
where $\digamma (t)=[\delta q(t),\delta p(t),\delta x(t),\delta y(t),\delta X_{1}(t),\delta
Y_{1}(t),\delta X_{2}(t)$,
$\delta Y_{2}(t)]^{T}$ is the is the vector of quadrature fluctuation
operators, and $\mathcal{N}(t)=[0,\xi (t),\sqrt{2k_{m}}x_{1}^{in}(t),\sqrt{%
2k_{m}}y_{m}^{in}(t)$,$\sqrt{2k_{1}}X_{1}^{in}(t),\sqrt{2k_{1}}Y_{1}^{in}(t),\sqrt{2k_{2}}%
(X_{2}^{in}(t),Y_{2}^{in}(t))]^{T}$  is the vector of input noises.
Furthermore, $\mathcal{M}$ is the drift matrix, expressed by:
\begin{equation*}
\mathcal{M}=\left(
\begin{array}{cccccccc}
0 & \omega _{b} & 0 & 0 & 0 & 0 & 0 & 0 \\
-\omega _{b} & -\gamma _{b} & 0 & G & 0 & 0 & 0 & 0 \\
-G & 0 & -\kappa _{m} & \Delta _{+} & 0 & \Gamma _{1} & 0 & \Gamma _{2} \\
0 & 0 & \Delta _{-} & -\kappa _{m} & -\Gamma _{1} & 0 & -\Gamma _{2} & 0 \\
0 & 0 & 0 & \Gamma _{1} & -\kappa _{1} & \Delta _{1} & 0 & 0 \\
0 & 0 & -\Gamma _{1} & 0 & -\Delta _{1} & -\kappa _{1} & 0 & 0 \\
0 & 0 & 0 & \Gamma _{2} & 0 & 0 & -\kappa _{2} & \Delta _{2} \\
0 & 0 & -\Gamma _{2} & 0 & 0 & 0 & -\Delta _{2} & -\kappa _{2}%
\end{array}%
\right) ,
\end{equation*}%
where $\Delta _{\pm }=\pm (\Delta _{m}+\Delta _{\mathcal{K}_{1}})-\Delta _{%
\mathcal{K}_{2}}$ and $G=i\sqrt{2}g_{mb}m_{s}$ is the effective
magnomechanical coupling rate.
\subsection{ENTANGLEMENT MEASURES}
The stability of the current magnomechanical system is the primary conditionthe
and the basic criteria for stability is the Routh-Hurwitz criterion \cite{RHCr,Shl3}. According to this criterion, the drift matrix must have a
negative real part of all the eigenvalues. Thus, using the secular
equation i.e., $|\mathcal{M}-\lambda_{\mathcal{M}} \mathbb{I}|=0$), we must
extract the eigenvalues from the drift matrix $\mathcal{M}$, and confirm the
system's stability. Our magnomechanical system's drift matrix is a $8\times 8
$ matrix, hence the corresponding covariance matrix will also be a $8\times 8
$ matrix, with the entries $\mathcal{V}_{ij}(t)=\frac{1}{2}\left\langle
\digamma_{i}(t)\digamma_{j}(t^{\prime})+\digamma_{j}(t^{\prime })\digamma_{i}(t)\right\rangle$, where the steady-state $\mathcal{V}$ can be
obtained by solving the steady-state Lyapunov equation \cite{Parks,SA},
\begin{equation}
\mathcal{M}\mathcal{V}+\mathcal{V}\mathcal{M}^{T}=-\mathit{D},  \label{f}
\end{equation}%
where $\mathit{D}=$ diag$[0,\gamma _{b}\left( 2n_{b}+1\right),\kappa
_{m}\left( 2n_{m}+1\right),\kappa _{m}( 2n_{m}+
1)$, $\kappa _{1}\left(2n_{1}+1\right),\kappa
_{1}\left(2n_{1}+1\right),\kappa _{2}\left(2n_{2}+1\right),\kappa
_{2}\left(2n_{2}+1\right)]$ is the diffusion
matrix, with $D_{ij}\delta (t-t^{\prime})=\frac{1}{2}\left\langle
\mathcal{N}_{i}(t)\mathcal{N}_{j}(t^{\prime})+\mathcal{N}_{j}(t^{%
\prime })\mathcal{N}_{i}(t)\right\rangle$.
We employ Simon's condition for
Gaussian states to simulate the bipartite entanglement \cite%
{Gonzalez,Vidal,Plenio,Adesso,SA, MCO1}.
\begin{equation}
E_{N}=\max [0,-\ln 2 (\nu ^{-})],  \label{LoN}
\end{equation}%
where $\nu^{-}=$min eig$|\bigoplus^{2}_{j=1}(-\sigma_{y})\widetilde{\mathcal{%
V}_{4}}|$ defines the minimum symplectic eigenvalue of the CM which is
reduced to an order of $4\times 4$. Here, $\widetilde{\mathcal{V}_{4}}%
=\varrho_{1|2}V_{in}\varrho_{1|2}$, where $V_{in}$ is a $4\times4$ matrix of
any bipartition. Furthermore, by removing the uninteresting columns and rows
in $\mathcal{V}_{4}$, we can generate $V_{in}$. The matrix $\varrho_{1|2}$%
=diag$(1,-1,1,1)$=$\sigma_{z}\bigoplus \mathbb{I}$, where $\sigma$'s are the
Pauli's spin matrices and $\mathbb{I}$ is the identity matrix, set the
partial transposition at the CM level. More intuitively, we introduce the
bipartite nonlinear index as
\begin{equation}
\Delta E_{K}=|E_{N}(\Delta_{K}>0)-E_{N}(\Delta_{K}<0)|,  \label{NLI}
\end{equation}
It is crucial to mention here that there will be nonreciprocity in bipartite entanglement if $\Delta E_{K}>0$, or $\Delta E_{K}\neq E_{N}(\Delta_{K}>0)$, or $\Delta E_{K}\neq E_{N}(\Delta_{K}>0)$.

To investigate the tripartite entanglement, we employ minimal residual
contangle, as given in \cite{Adesso,Adesso2,Coffman}
\begin{equation}
R_{\alpha-\beta-\gamma}^{\tau-min}\equiv \min [\mathcal{R}^{\tau}_{\alpha|\beta\gamma},\mathcal{R}^{\tau}_{\beta | \alpha\gamma},\mathcal{R}^{\tau }_{\gamma|\alpha\beta}],  \label{RS}
\end{equation}%
where $\mathcal{R}^{\tau}_{\alpha|\beta\gamma}$ assure the invariance of tripartite
entanglement under all permutations of the modes, and is given by
\begin{equation}
\mathcal{R}^{\tau }_{\alpha|\beta\gamma}\equiv C_{\alpha|\beta\gamma}-C_{\alpha|\beta}-C_{\alpha|\gamma}\geq 0, \ \
(\alpha,\beta,\gamma=p,m,b).
\end{equation}
$C_{\alpha|\beta}$ is the proper entanglement monotone and is equal to the square of
logarithmic negativity of the subsystems, i.e., $C_{\alpha|\beta}=E_{\alpha|\beta}^{2}$, where
$\alpha$ incorporates one or two modes. However, to compute the \textit{%
one-mode-vs-two-modes} logarithmic negativity $E_{\alpha|\beta\gamma}$, one must redefine
the definition of $\nu^{-}$ as given by 
\begin{equation}
\eta ^{-}=\min eig\ |\oplus_{j=1}^{3}(-\sigma _{y})\widetilde{V_{j}}|,
\end{equation}
where $\widetilde{V_{j}}=\mathcal{P}_{i|jk}\mathcal{V}_{6}\mathcal{P}_{i|jk}$
($i \neq j \neq k $) with $\mathcal{V}_{6}$ being the $6\times6$ CM of the
three modes of interest. Furthermore, $\mathcal{P}_{1|23}=\sigma_{z}\oplus
\mathbb{I}\oplus \mathbb{I}$, $\mathcal{P}_{2|13}= \mathbb{I}\oplus
\sigma_{z}\oplus \mathbb{I}$ and $\mathcal{P}_{3|12}=\mathbb{I}\oplus\mathbb{%
I}\oplus \sigma_{z}$, are the matrices that define the partial transposition
at the level of the CM $\mathcal{V}_{6}$. In addition, $\sigma
_{y}=[0,-i;i,0]$ and $\sigma _{z}=[1,0;0,-1]$. Moreover, the symbol $\oplus$%
, which describes the direct sum of the matrices, can expand the dimension to
$(m+p)\times(n+q)$ of two matrices $A$ with $m\times n$ and $B$ with $%
p\times q$ dimension.

More intuitively, we also introduce the bidirectional contrast ratio for tripartite entanglement as
\begin{equation}
B_{\alpha-\beta-\gamma}^{cr}=\frac{|R_{\alpha-\beta-\gamma}^{\tau-min}(\Delta_{K}>0)-R_{\alpha-\beta-\gamma}^{\tau-min}(\Delta_{K}<0)|}{R_{\alpha-\beta-\gamma}^{\tau-min}(\Delta_{K}>0)+R_{\alpha-\beta-\gamma}^{\tau-min}(\Delta_{K}<0)}.
\label{NLITR}
\end{equation}
There exists no (ideal) nonreciprocity in tripartite entanglement if $B_{\alpha-\beta-\gamma}^{cr}=0$ ($B_{\alpha-\beta-\gamma}^{cr}=1$). The Stronger nonreciprocity of tripartite entanglement is indicated by a larger value of the bidirectional contrast ratio $B_{\alpha-\beta-\gamma}^{cr}$.
\begin{figure}[b!]
\centering
\includegraphics[width=0.95\columnwidth,height=3in]{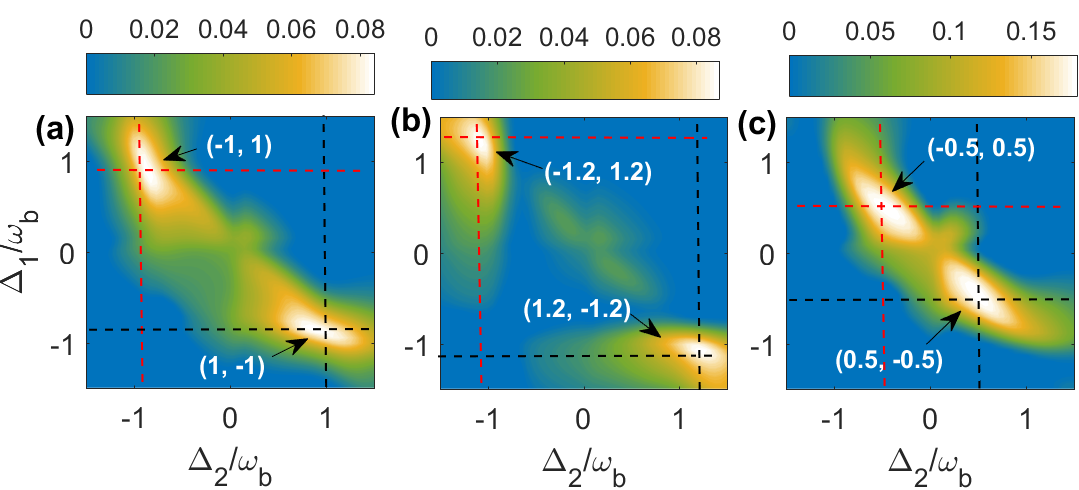}
\caption{Contour plot of $E_{N}^{c1-c2}$ versus the normalized cavity
detunings $\Delta _{1}/\protect\omega _{b}$ and $\Delta _{2}/\protect\omega %
_{b}$ when (a) $\Delta_{K}=0$ (b) $\Delta_{K}>0$ (c) $\Delta_{K}<0$. The
other parameters are listed in the main text.}
\end{figure}
\begin{figure}[b!]
\centering
\includegraphics[width=0.95\columnwidth,height=3in]{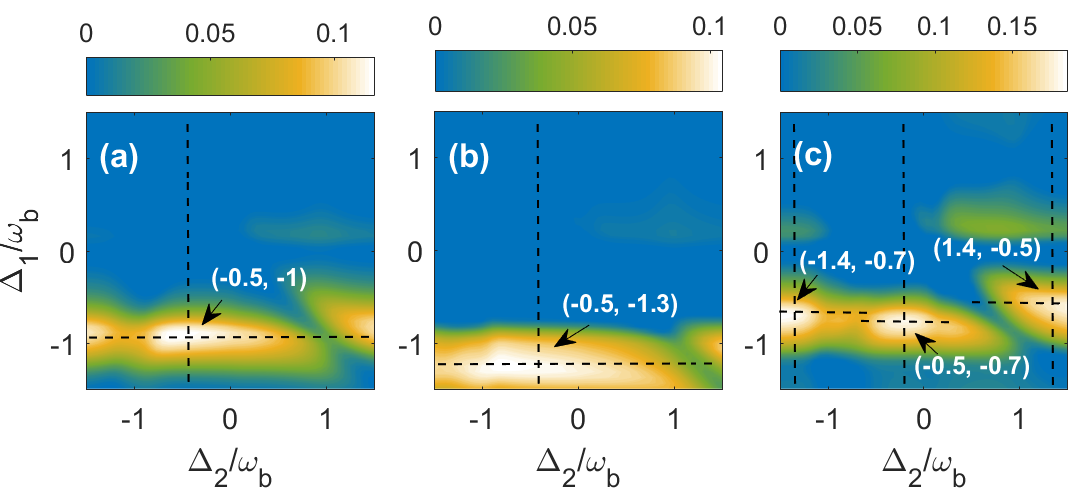} \centering
\caption{Contour plot of $E_{N}^{c1-m}$ versus the normalized cavity
detunings $\Delta _{1}/\protect\omega _{b}$ and $\Delta _{2}/\protect\omega %
_{b}$ when (a) $\Delta_{K}=0$ (b) $\Delta_{K}>0$ (c) $\Delta_{K}<0$. The
other parameters are listed in the main text.}
\end{figure}
\section{Kerr-INDUCED NONRECIPROCAL Multipartite ENTANGLEMENT}
At the outset of the discussion, we should mention here that the nonreciprocal
entanglement due to the magnon Kerr effect is quite different from the case
of the Sagnac effect. This is because the magnetic
field-mediated Kerr effect generates either a red or blue shift in magnon
frequency. Furthermore, the primary goal of investigating entanglement in
such a dual-cavity magnomechanical system with and without Kerr
nonlinearity is to find optimal detunings among the four modes to establish
true bipartite and tripartite entanglement. To study nonreciprocal
entanglement, we choose the experimentally feasible parameters: $%
\omega_{1}=\omega_{2}=\omega=2\pi\times 10$ GHz, $\omega_{b}=2\pi \times 10$ MHz, $\kappa_{1}=\kappa_{2}=\kappa=0.1\omega_{b}$ MHz, $\kappa_{m}=0.2\omega_{b}$,
$\Gamma_{1}=\Gamma_{2}=\Gamma=0.32\omega_{b}$, $\gamma_{b}=10^{-5}\omega_{b}$, $g_{mb}=2\pi \times 0.3$ Hz, $T=10$ mK, $H_d=1.3\times10^-{4}, \gamma_{G}/2\pi=28$GHz/T, $r=50\mu$m, $\rho=4.22\times10^{27}$ and $P_j=50$mW \cite{Non1,Non2}. We mainly
consider four bipartitions, namely, cavity-1 photon-cavity-2 photon $E_{N}^{c1-c2}$,
cavity-1 photon-magnon $E_{N}^{c1-m}$, cavity-1 photon-phonon $E_{N}^{c1-b}$ and
magnon-phonon $E_{N}^{m-b}$.
In addition, we also consider two tri-partition, namely, photon-magnon-phonon $\mathcal{R}^{\tau-min}_{c1-m-b}$ and photon-magnon-photon $\mathcal{R}^{\tau-min}_{c1-m-c2}$.
The subfigures (b) and (c) in Fig. (2) to Fig.
(5) respectively indicate the magnetic field along the crystallographic axis
[100] and [110], which correspond to $\Delta_{K}>0$ and $\Delta_{K}<0$.
However, for comparison, the subfigures (a) in Fig. (2) to Fig. (5) correspond to
the case of entanglement without Kerr nonlinearity, i.e., $\Delta_{K}=0$. The value of Kerr-nonlinearity
may vary, depending upon the magnon number.
This is because the parameters for both cavities are similar and magnon/phonon
entanglements with either cavity 1 or cavity 2 are identical. It is worth
mentioning that by incorporating Kerr nonlinearity, the value of the effective
magnomechanical coupling strength is different from the case without Kerr
nonlinearity.
The value of the effective magnomechanical coupling strength is $G/\Gamma=1.1
$, $G/\Gamma\approx1$ and $G/\Gamma\approx1.4$ for $\Delta_{K}=0$, $%
\Delta_{K}>0$, $\Delta_{K}<0$, respectively. In addition, the chosen
parameters to ensure the stability of the system in accordance with the
Routh-Hurwitz criterion \cite{RHCr}.
\begin{figure}[tbp]
\centering
\includegraphics[width=0.95\columnwidth,height=3in]{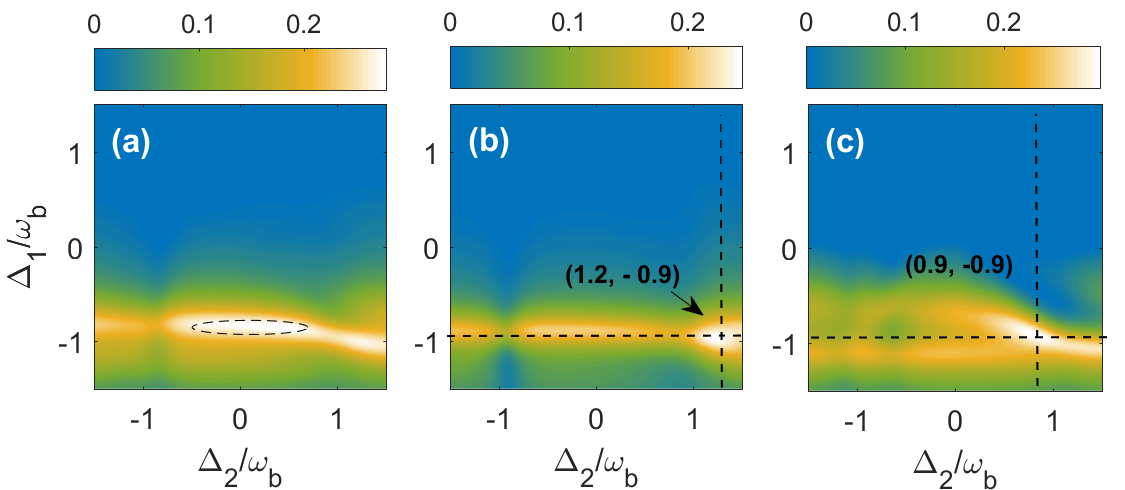} \centering
\caption{Contour plot of $E_{N}^{c1-b}$ versus the normalized cavity
detunings $\Delta _{1}/\protect\omega _{b}$ and $\Delta _{2}/\protect\omega %
_{b}$ when (a) $\Delta_{K}=0$ (b) $\Delta_{K}>0$ (c) $\Delta_{K}<0$. The
other parameters are listed in the main text.}
\end{figure}
\begin{figure}[tbp]
\centering
\includegraphics[width=0.95\columnwidth,height=3in]{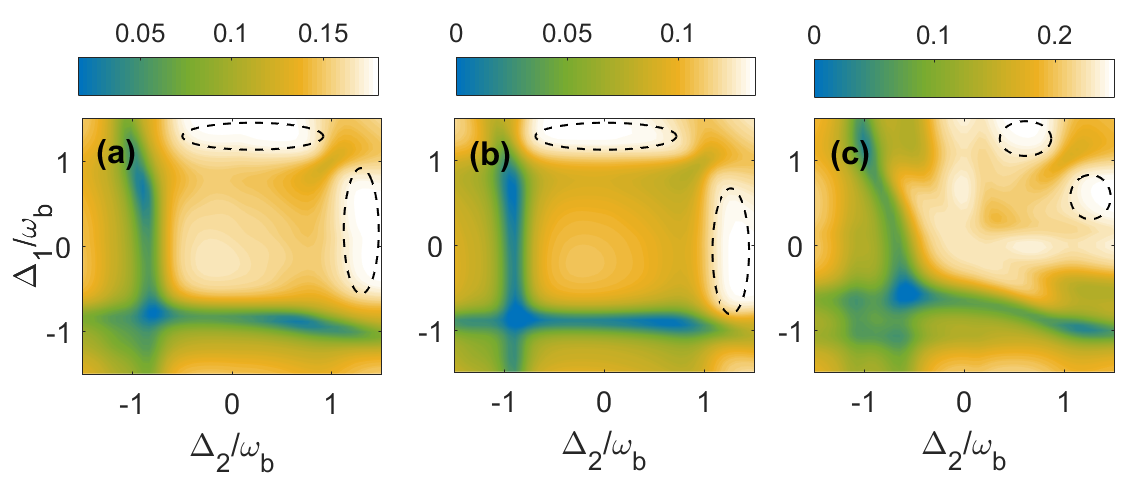} \centering
\caption{Contour plot of $E_{N}^{m-b}$ versus the normalized cavity
detunings $\Delta _{1}/\protect\omega _{b}$ and $\Delta _{2}/\protect\omega %
_{b}$ when (a) $\Delta_{K}=0$ (b) $\Delta_{K}>0$ (c) $\Delta_{K}<0$. The
other parameters are listed in the main text.}
\end{figure}

\subsection{Bipartite Entanglement}
To study the effects of Kerr non-linearity on the bipartite entanglement, we have obtained several results. These results incorporate the dynamics of nonreciprocal entanglement in the absence and presence of Kerr nonlinearity, with the direction of magnon squeezing on the different bipartite entanglement, as shown in Figs. 2, 3, 4, and 5. First, we discuss the
contour plot of the photon-photon entanglement, $E_{N}^{c1-c2}$, as function of
the normalized detunings $\Delta_{1}/\omega_{b}$ and $\Delta_{2}/\omega_{b}$%
, respectively, keeping the magnon detuning $\Delta_{m}/\omega_{b}=1$, as shown
in Fig. 2 (a-c). A careful analysis of these results shows that the
generated entanglement is nonreciprocal and strongly depends on thethe
value of self-Kerr coefficient, $\Delta_{K}$ .
It is clear that in the
absence of Magnon self-Kerr effect, entanglement initially has maximum
values around $\Delta_{2}/\omega_{b}= \pm 1$. However, in the presence of
self-Kerr effect, the maxima of entanglement is not only enhanced in the
magnitude but also shifted towards higher (lower) detuning region when $%
\Delta_{K}>0$ ($\Delta_{K}<0$), showing the non-reciprocity of generated
entanglement. This supports our claim that the magnon nonlinear self-Kerr effect has a noticeable effect on the entanglement. This
provides additional control for the manipulation of cavity-cavity bipartite entanglement.

Next, in Fig. 3, we show the results for the bipartite entanglement
between cavity photons and magnons, for three different choices of $\Delta_{K}$. It can be seen that the entanglement in the absence of self-Kerr
nonlinearity is different from the results after the inclusion of
nonlinearity. It once again shows the nonreciprocal nature of
entanglement for the same system parameters used in Fig. 2. In a similar
line of action, we also studied the effects of self-Kerr nonlinearity on the
cavity photon-phonons (Fig. 4) and magnon-phonons (Fig. 5) bipartitions.
It can be seen that all bipartite entanglement is enhanced and has been shifted, compared to the results for the absence of nonlinearity.
It is worth mentioning here that the nonreciprocal entanglement is because
of the application of strong drive fields and then magnon self-Kerr
nonlinearity gives rise to a shift in magnon frequency. The direction of the applied magnetic field is responsible for the sign of the induced frequency
shift. Hence, considering the crystallographic rotation effect of the magnon mode, the magnon significantly enhances the bipartite entanglements between different modes more than the previous models which don't have a crystallographic rotation effect inspite of having kerr-nonlinearity \cite{IR3,yang2020entanglement}.

In reality, the magnomechanical coupling strength can be modulated by varying
the input driving fields [see Eq. (5)] in our approach. Therefore, it is
crucial to observe the effect of the magnomechanical coupling with or without
the magnon-Kerr effect on different bipartitions. We first choose the value of
cavity detunings $\Delta_{1}$ and $\Delta_{2}$ where we found the optimal
entanglement without Kerr nonlinearity and then observe how Kerr
nonlinearity affects the effective coupling rate i.e. $G$. So, we
plot all bipartitions against the effective coupling rate $G$, in the absence/presence of the magnon self-Kerr nonlinearity. Fig. 6(a-d)  shows the plot for entanglement in all four bipartitions for different choices of Kerrthe
nonlinearities, i.e., $\Delta_{K}=0$, $\Delta_{K}>0$, $\Delta_{K}<0$,
respectively. An additional quantity $\Delta E_{K}$, defined in Eq. (9) as the
bipartite non-linear index (yellow curve), represents the reciprocity in the system.
This result clearly shows that an increase in the effective coupling rate $G$
results in an increase in entanglement. The nonreciprocity of entanglement is
self-evident from this plot
and can also be observed by the yellow curve,
bipartite nonlinear index $\Delta E_{K}$. When $\Delta E_{K}=0$ or overlaps, the red/blue curve represents no reciprocity.
In Fig. 6(e-h), we plot the entanglement of all bipartitions and observe the simultaneous nonreciprocity when $\Delta_{1}/\omega_{b}=-1$ and $\Delta_{2}/\omega_{b}=+1$.
These results show that the smaller value of effective coupling rate $G$ results in optimal entanglement for $\Delta E_{K}<0$ while the larger value of $G$ yields optimal entanglement for $\Delta E_{K}>0$, exhibits the presence of nonreciprocal entanglement of all bipartitions and the enhancement of entanglement for nonzero self-Kerr nonlinearity. Based on the results obtained, we can assert that the nonreciprocal entanglement can be manipulated and enhanced for required applications, using the optimal system parameters.
\begin{figure}[tbp]
\centering
\includegraphics[width=0.98\columnwidth,height=3.3in]{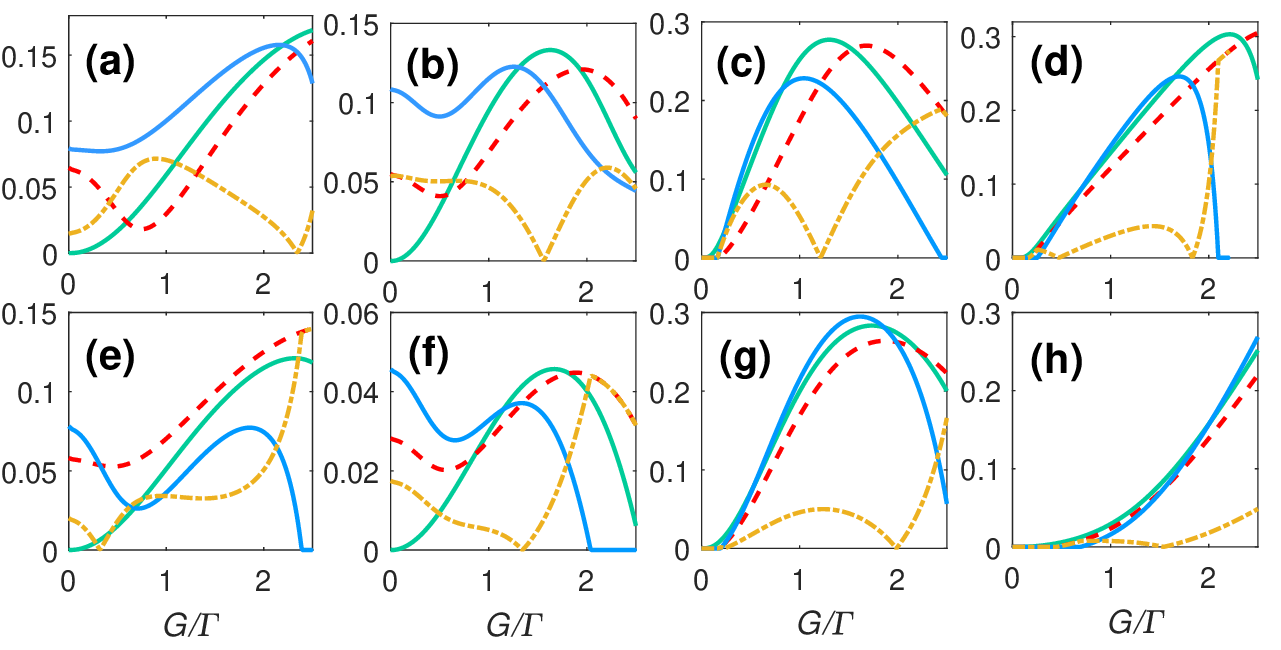} \centering
\caption{Plot of (a)(e) $E_{N}^{c1-c2}$, (b)(f) $E_{N}^{c1-m}$, (c)(g) $E_{N}^{c1-b}$,
and (d)(h) $E_{N}^{m-b}$ versus the ratio of $G/\Gamma$ with $\Delta_{K}=0$
(the green curve), $\Delta_{K}>0$ (the red curve), $\Delta_{K}<0$ (the blue
curve), and $\Delta E_{K}$ (the yellow curve). In (a-d), we choose optimum value of entanglements when $\Delta_{K}=0$ as the axis depicted from the first subfigures of Fig. (2)-Fig. (5), while (e-f), we take the optimum values of entanglements when $\Delta_{1}=-1$ and $\Delta_{2}=1$ }
\end{figure}
\begin{figure}[tbp]
\centering
\includegraphics[width=0.98\columnwidth,height=4.45in]{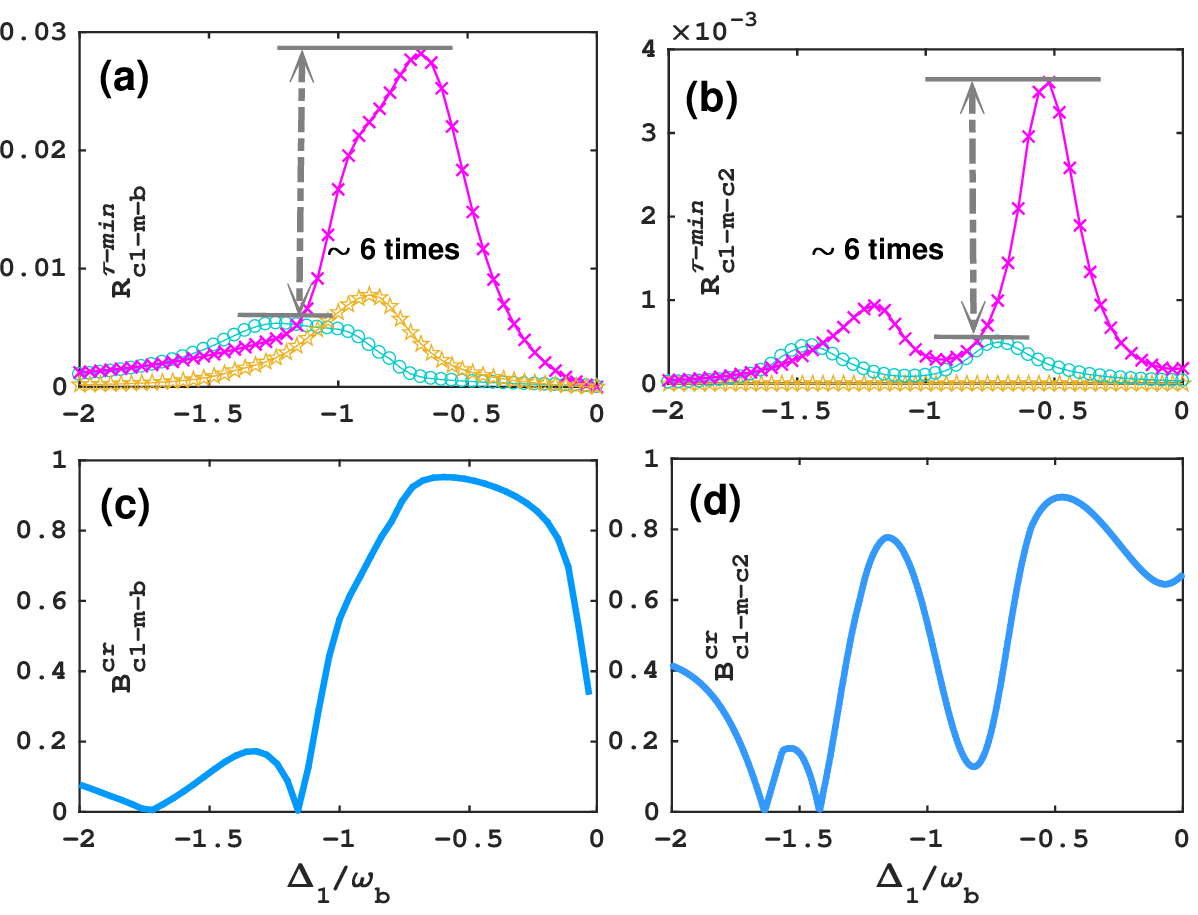} \centering
\caption{Tripartite entanglement in terms of the minimum residual contangle
\textcolor{blue}{(a) $\mathcal{R}^{\tau-min}_{c1-m-b}$ and (b) $\mathcal{R}^{\tau-min}_{c1-m-c2}$} versus cavity detuning $\Delta_{1}/\protect\omega_{b}$
with $\Delta_{K}=0$ (yellow curve), $\Delta_{K}>0$ (green curve), and $\Delta_{K}<0$ (magenta curve). Bidirectional contrast ratio (c) $B^{cr}_{c1-m-b}$ (d) $B^{cr}_{c1-m-c2}$
for tripartite entanglements
as functions of $\Delta_{1}/\omega _{b}$.}
\end{figure}
\subsection{Tripartite Entanglement}
Apart from Kerr-dependent bipartite entanglement, the nonreciprocal
photon-magnon-phonon $\mathcal{R}^{\tau-min}_{c1-m-b}$ and
photon-magnon-photon $\mathcal{R}^{\tau-min}_{c1-m-c2}$ tripartite
entanglements could be obtained. In addition, the nonreciprocity of these
tripartite entanglement could be enhanced by Kerr effects, as expected, and are
numerically illustrated in Fig. 7(a-b). It is obvious from Fig. 7(a) that the
tripartite entanglement $\mathcal{R}^{\tau-min}_{c1-m-b}$ curve shifts to the
left (for $\Delta_{k}> 0$) or right (for $\Delta_{k}< 0$), however, we
obtained enhanced tripartite entanglement $\mathcal{R}^{\tau-min}_{c1-m-b}$
when $\Delta_{k}< 0$. Furthermore, there exist photon-magnon-photon $\mathcal{R}^{\tau-min}_{c1-m-c2}$ tripartite entanglement only when $%
\Delta_{k}> 0$ or $\Delta_{k}< 0$.
Based on Eq. (\ref{NLITR}), it is clear that the higher the value of bidirectional contrast ratio $B_{\alpha-\beta-\gamma}^{cr}$, the
stronger nonreciprocity of the tripartite entanglement is. In Fig. 7(c-d), we numerically exhibit the bidirectional contrast ratios $B^{cr}_{c1-m-b}$ and $B^{cr}_{c1-m-c2}$ against $\Delta_{1}$ to make this point very evidently. We note that the nonreciprocal tripartite entanglement manifestly appears when the bidirectional contrast ratio for tripartite entanglement is nonzero, however, the nonreciprocity vanishes when the bidirectional contrast ratio for tripartite entanglement becomes zero as shown by Fig. 7(c-d).
Hence, by changing $\Delta_{1}$, the bidirectional contrast ratios for the tripartite entanglement can be adjusted from 0 to 1. This suggests that genuine tripartite entanglements with almost ideal nonreciprocity can be acquired in our scheme, via tuning the cavity frequency detuning.
\subsection{Feasibility of Current Scheme}
In this section, we discuss the experimental feasibility of effective
squeezing caused by self-Kerr nonlinearity. Note that the coefficient of
Kerr nonlinearity is inversely proportional to the volume of the YIG sphere,
so the Kerr nonlinear effect can become significant when we use a YIG
sphere of small size \cite{Non1}. The magnon excitation number for magnetic materials must be much lower than to guarantee the validity of the magnon description. For a 250$\mu$m diameter YIG sphere, $\left\vert m_{s} \right\vert ^{2}
\simeq 10^{14}\ll 5N \simeq 10^{17}$ \cite{mag}, which meets the
low-excitation condition, and guarantees that the Kerr nonlinearity is
considerably small. In addition, a YIG of 250$\mu$m diameter generates Kerr
coefficient $\mathcal{K}_{r}=6.4*10^{-9}$ and, to keep the Kerr effect
negligible, $\Omega\gg\mathcal{K}_{r}\left\vert m_{s} \right\vert ^{3}$ must be fulfilled. Therefore, $\Omega= 7.1\times10^{14}  \gg\mathcal{K}_{r}\left\vert
m_{s} \right\vert ^{3} = 5.7\times10^{13}$. It is vital to note that the
self-Kerr nonlinearity cannot be ignored if $\mathcal{K}_{r}\left\vert m_{s}
\right\vert ^{3} \gtrapprox \Omega$. However, we use a $0.1$mm-diameter YIG sphere to generate a strong Kerr coefficient $\mathcal{K}_{r}\approx
\times10^{-7}$ \cite{Theor} and therefore,  $\Omega= 7.5\times10^{14}\approx
\mathcal{K}_{r}\left\vert m_{s} \right\vert ^{3} = 7.8\times10^{14}$. Hence,
the experimental values show the validity of parameter regimes considered in the
current system and the above condition makes our magnomechanical system with Kerr-nonlinearity sufficiently practicable.
In terms of experimental realization, a planar cross-shaped cavity or coplanar waveguide can be used to implement the current method.

\section{Conclusion}
In this study, we have theoretically proposed an efficient proposal for generating nonreciprocal entanglement in a two-mode magnomechanical system. We exploited the magnon self-Kerr nonlinear effect, which leads to enhanced nonreciprocal entanglement. We have shown that the degree of nonreciprocity is the manipulation of optimal choice of system parameters like normalized cavity detunings, bipartite nonlinear index $\Delta E_{K}$, self-Kerr coefficient, and effective magnomechanical coupling rate G. It is worth mentioning that Kerr nonlinearity gives rise to a shift in magnon frequency by the application of a strong drive field. The direction of the applied magnetic field is responsible for the sign of the induced frequency shift. This results in nonreciprocal bipartite (tripartite) entanglement, which can be seen by the bipartite nonlinear index (bidirectional contrast ratio). The results were presented for both bipartite (all possible bipartitions) and tripartite entanglements by utilizing experimentally feasible parameters. In the end, we can safely assert that the present proposal may find some practical applications in nonreciprocal device engineering and nonlinear cavity magnomechanical system applications with nonlinear effects.
\section*{Declaration of interest}
The authors declare that they have no known competing financial interests or
personal relationships that could have appeared to influence the work
reported in this paper.
\section*{Data availability}
All numerical data that support the findings in this study is available within the article.

\section*{Acknowledgement}
Amjad Sohail and Marcos C\'{e}sar de Oliveira gratefully acknowledges the financial support given by CNPq through grant number 171707/2023-0.
S. K. Singh gratefully acknowledges the financial support given by Universiti
Teknologi Malaysia through grant number Q.J130000.21A2.07E14.

\end{document}